\documentclass[12pt]{article}
\usepackage{epsf, cite, amssymb}
\usepackage{epsfig}
\usepackage{amsmath}
\usepackage{latexsym}
\usepackage{epsfig}
\usepackage{cite}
\usepackage[polish,english]{babel}
\usepackage{graphicx,color}
\usepackage{epstopdf}
\usepackage[colorlinks=true, linkcolor=blue, bookmarks=true, citecolor=violet]{hyperref}

\newcommand{\arXiv}[1]{\href{http://www.arXiv.org/abs/#1}{#1}}

\makeatletter
\renewcommand\section{\@startsection {section}{1}{\z@}%
                               {-3.5ex \@plus -1ex \@minus -.2ex}
                               {2.3ex \@plus.2ex}%
                               {\normalfont\large\bfseries}}
\renewcommand\subsection{\@startsection{subsection}{2}{\z@}%
                                 {-3.25ex\@plus -1ex \@minus -.2ex}%
                                 {1.5ex \@plus .2ex}%
                                 {\normalfont\bfseries}}
\makeatother

\usepackage{xcolor}

\def\IZ{\relax\ifmmode\mathchoice
{\hbox{\cmss Z\kern-.4em Z}}{\hbox{\cmss Z\kern-.4em Z}}
{\lower.9pt\hbox{\cmsss Z\kern-.4em Z}} {\lower1.2pt\hbox{\cmsss
Z\kern-.4em Z}}\else{\cmss Z\kern-.4em Z}\fi}
\def\IR{\relax{\rm I\kern-.18em R}}

\def\one{{\hbox{ 1\kern-.8mm l}}}

\newlength{\bredde}
\def\slash#1{\settowidth{\bredde}{$#1$}\ifmmode\,\raisebox{.15ex}{/}
\hspace*{-\bredde} #1\else$\,\raisebox{.15ex}{/}\hspace*{-\bredde}
#1$\fi}

\newsavebox{\zzzbar}
\sbox{\zzzbar}
{\setlength{\unitlength}{0.9em}
\begin{picture}(0.6,0.7)
\thinlines
\put(0,0){\line(1,0){0.6}}
\put(0,0.75){\line(1,0){0.575}}
\multiput(0,0)(0.0125,0.025){30}{\rule{0.3pt}{0.3pt}}
\multiput(0.2,0)(0.0125,0.025){30}{\rule{0.3pt}{0.3pt}}
\put(0,0.75){\line(0,-1){0.15}}
\put(0.015,0.75){\line(0,-1){0.1}}
\put(0.03,0.75){\line(0,-1){0.075}}
\put(0.045,0.75){\line(0,-1){0.05}}
\put(0.05,0.75){\line(0,-1){0.025}}
\put(0.6,0){\line(0,1){0.15}}
\put(0.585,0){\line(0,1){0.1}}
\put(0.57,0){\line(0,1){0.075}}
\put(0.555,0){\line(0,1){0.05}}
\put(0.55,0){\line(0,1){0.025}}
\end{picture}}

\def\Re{{\rm Re ~}}

\newcommand{\ena}{\end{eqnarray}}
\newcommand{\beqa}{\begin{eqnarray}}
\newcommand{\eeqa}{\end{eqnarray}}
\newcommand{\bea}{\begin{eqnarray}}
\newcommand{\eea}{\end{eqnarray}}

\newcommand{\be}{\begin{equation}}
\newcommand{\ee}{\end{equation}}

\usepackage{graphicx}

\def\be{\begin{equation}}
\def\ee{\end{equation}}
\def\beq{\begin{eqnarray}}
\def\eeq{\end{eqnarray}}

\def\p{\partial}

\def\({\left (}
\def\){\right )}
\def\[{\left [}
\def\[{\right ]}

\def\ba{\begin{eqnarray}}
\def\ea{\end{eqnarray}}

\def\p{\partial}

\def \p{\partial}

\input amssym.def
\input amssym.tex

\newcommand{\bbibitem}[1]{\bibitem{#1}\marginpar{#1}}
\def\Bibitem#1{\bibitem{#1}%
  \smash{\hbox to0pt{\raise1ex\hbox{\tiny[#1]}\hss}}}

\def\Label#1{\label{#1}%
  \smash{\hbox to0pt{\raise1ex\hbox{\tiny[#1]}\hss}}}
\def\noLabels{\let\Label=\label}
\def\nobbibitem{\let\bbibitem=\bibitem}
 \def\noBibitem{\let\Bibitem=\bibitem}

\newcommand{\beann}{\begin{eqnarray*}}  \newcommand{\eeann}{\end{eqnarray*}}
\newcommand{\bfig}{\begin{figure}} \newcommand{\efig}{\end{figure}}
\newcommand{\bcen}{\begin{center}} \newcommand{\ecen}{\end{center}}
\newcommand{\btab}{\begin{tabular}} \newcommand{\etab}{\end{tabular}}

\renewcommand{\Re}{\mathop{\rm Re}}   
\newtheorem{Proposition}{Proposition}[section]

\newtheorem{Theorem}{Theorem}[section]
\newtheorem{Lemma}{Lemma}[section]
\newtheorem{Corrolary}{Corrolary}[section]

\newcommand{\bp}{\begin{Proposition}}   \newcommand{\ep}{\end{Proposition}}
\newcommand{\bt}{\begin{Theorem}}   \newcommand{\et}{\end{Theorem}}
\newcommand{\bl}{\begin{Lemma}}     \newcommand{\el}{\end{Lemma}}
\newcommand{\bc}{\begin{Corrolary}} \newcommand{\ec}{\end{Corrolary}}
\begin{document}

\begin{titlepage}
\begin{flushright}
IFJPAN-IV-2013-16
\end{flushright}
\vfill
\begin{center}
{\Large \bf Non-linear evolution of unintegrated gluon density at large values of coupling constant}

\vskip 10mm

{\large Krzysztof Kutak$^{a}$, Piotr Sur\'owka$^{b}$}

\vskip 7mm

$^a$Instytut Fizyki Jadrowej im H. Niewodnicza\'nskiego,\\
\hspace*{0.15cm} Radzikowskiego 152, 31-342 Krak\'ow, Poland\\

$^b$  Theoretische Natuurkunde, Vrije Universiteit Brussel, \\
\hspace*{0.15cm}  and International Solvay Institutes, \\
\hspace*{0.15cm} Pleinlaan 2, B-1050 Brussels, Belgium. \\

\vskip 3mm
\vskip 3mm
{\small\noindent  {\tt Krzysztof.Kutak@ifj.edu.pl,\\ Piotr.Surowka@vub.ac.be}}

\end{center}
\vfill

\begin{center}
{\bf ABSTRACT}
\vspace{3mm}
\end{center}
We propose an evolution equation for unintegrated gluon densities that is valid for large values of the QCD coupling constant $\bar{\alpha} _s$. Our approach is based on the linear resummation model introduced by Sta\'{s}to. We generalize the model including a non-linear term in the diffusive regime. The validity of the diffusive evolution at strong coupling is supported by the AdS/CFT consideration, as well as perturbative arguments. We solve the evolution equation numerically and extract the saturation scale, which we compare with the weak coupling counterpart.
\vfill


\end{titlepage}


\section{Introduction}
A strongly coupled quantum field theory like quantum chromodynamics is the basic framework which is used in the interpretation of hadronic observables data from the high-energy physics experiments. Despite its correctness many open questions remain, and the full theoretical description is far from complete, as neither perturbative methods nor lattice gauge theory can provide a full description of hadronic phenomena. One of the open problems is, for instance, gluon saturation \cite{Gribov:1984tu}, which is expected on theoretical grounds and there is growing evidence that it occurs \cite{Stasto:2000er,Albacete:2010pg,Dumitru:2010iy}. Another open problem is the derivation of the dynamics of strongly coupled systems, such as quark-gluon plasma, directly from a QCD Lagrangian.

When the energy is high enough, quarks and gluons are elementary degrees of freedom in QCD. Therefore, an essential ingredient to understand the collisions is the parton content of the hadrons that are being collided. At present, we do not have analytic methods to derive parton distribution functions. We can either use the perturbation theory for carefully chosen observables and resum ``infrared'' and ``collinear'' logarithms or use some simplified holographic model under analytic control.

A particularly interesting resummation approach, offering a possible although not definite interpretation of low $x$ data from the electron-proton collider HERA at Deutsches Elektronen-Synchrotron, was developed by Balitsky, Fadin, Kuraev, and Lipatov (BFKL) \cite{Kuraev:1976ge,Kuraev:1977fs,Balitsky:1978ic}. The idea is that the scattering process occurs through the exchange of the so-called Reggeized gluons. The two interacting Reggeized gluons are known in the literature as the Pomeron. These are effective particles emerging after resummation of $(\alpha _s \ln \frac{1}{x})^n$.  Such a procedure gives an evolution equation for unintegrated gluon distribution functions schematically written as
\be
\frac{\p f(x,k^2)}{\p \ln x_0/x}=K \otimes f(x,k^2),
\ee
where $K$ is the evolution kernel and $\otimes$ denotes a convolution with respect to transverse momenta. The main prediction of the BFKL evolution is given by the hadronic cross section of the form
\be
\sigma \propto s^{\alpha _P},
\ee
where $\alpha _P$ is known as the intercept.
The phenomenologically interesting regime of the applicability of the leading-order (LO) and next-to-leading-order (NLO) BFKL equation is limited to the weak coupling physics.
However, there is an ongoing activity based on integrability properties and connections to string theory to extend the BFKL eqvolution equation to any value of the coupling constant and therefore to provide resummation of a large fraction of terms in the perturbative series (see\cite{Janik:2013nqa} and references therein).

The BFKL leads to powerlike growth of the gluon density with energy. This is a consequence of the violation of unitarity by the BFKL equation. The point at which the linear BFKL formalism has to be corrected to include non-linear effects is known as the saturation scale. Several approaches were proposed to encapsulate parton saturation effects such as recombination or rescattering \cite{Gribov:1984tu,Mueller:1985wy,McLerran:1993ni,Balitsky:1995ub,Kovchegov:1999yj,Kovchegov:1999ua,GolecBiernat:1998js,Ayala:1996em,AyalaFilho:1997du,JalilianMarian:1997gr,Iancu:2000hn,Kutak:2011fu,Kutak:2012yr,Kutak:2012qk}. The common feature of these approaches is that they introduce a non-linearity that takes into account saturation effects. So far no method based on integrability has been applied to shed light on saturation physics.

The LO BFKL equation improved with effects modelling higher-order terms, as observed in Ref. \cite{Stasto:2007uv}, can be useful in studies of infinite strong coupling effects. It has been noted that with the appropriate introduction of Dokshitzer-Gribov-Lipatov-Altarelli-Parisi (DGLAP) anomalous dimension into the BFKL framework, and, after resummation of kinematical effects to infinite order, one is able to extend formally the solution of the BFKL equation to large values of the coupling constant. One can ask whether at the strong coupling where due to large corrections from resummation the saturation effects will be necessary, i.e., whether gluon density will diverge for low values of the gluon's transversal momentum. In this paper, using the framework developed by Sta\'{s}to, which allows one to obtain gluon density in the large $\bar{\alpha} _s$ ($\bar{\alpha} _s\equiv N_c\alpha_s/\pi$) regime, we conclude that it is not enough to use the linear BFKL approach enhanced with kinematical corrections and DGLAP effects to provide a not diverging gluon density even at large coupling. As a way out to extend the approach of Ref. \cite{Stasto:2007uv}, we propose to use an appropriately resummed nonlinear Balitsky-Kovchegov equation.
Our results are obviously model dependent; however, they point at a potential problem with investigations based on entirely linear equations.\\
The paper is organized as follows. In Sec. \ref{BFKLrevised}, we review the basics of the BFKL equation. In Sec. \ref{BFKLstrong}, we solve the BFKL equation in the large values of $\bar{\alpha}_s$ and argue that the gluon distribution continues to evolve in a diffusive way. In Sec. \ref{AdSpomeron}, we further motivate the use of the diffusion approximation using results derived in holography. Finally in Sec. \ref{nonlineraeq}, we show how the non-linear Balitsky-Kovchegov equation can be applied to study the dynamics of unintegrated gluon densities at large values of $\bar{\alpha} _s$. Moreover, we extract a saturation scale at strong coupling which is qualitatively similar to the result obtained previously from gauge and gravity duality.

\section{The BFKL equation in diffusion approximation} \label{BFKLrevised}
\begin{figure}[t!]
  \begin{picture}(30,30)
    \put(20, -80){
      \includegraphics{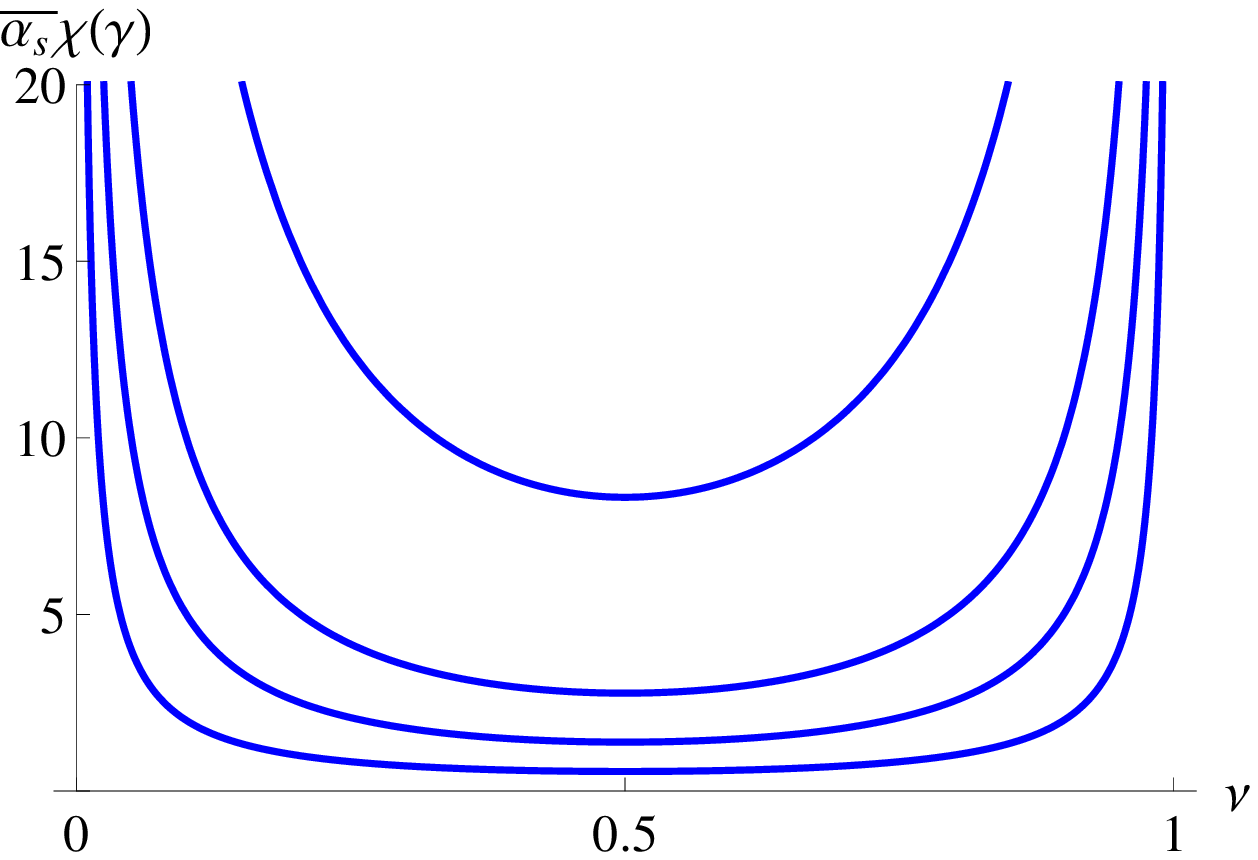}
    }

\put(280, -80){
      \includegraphics{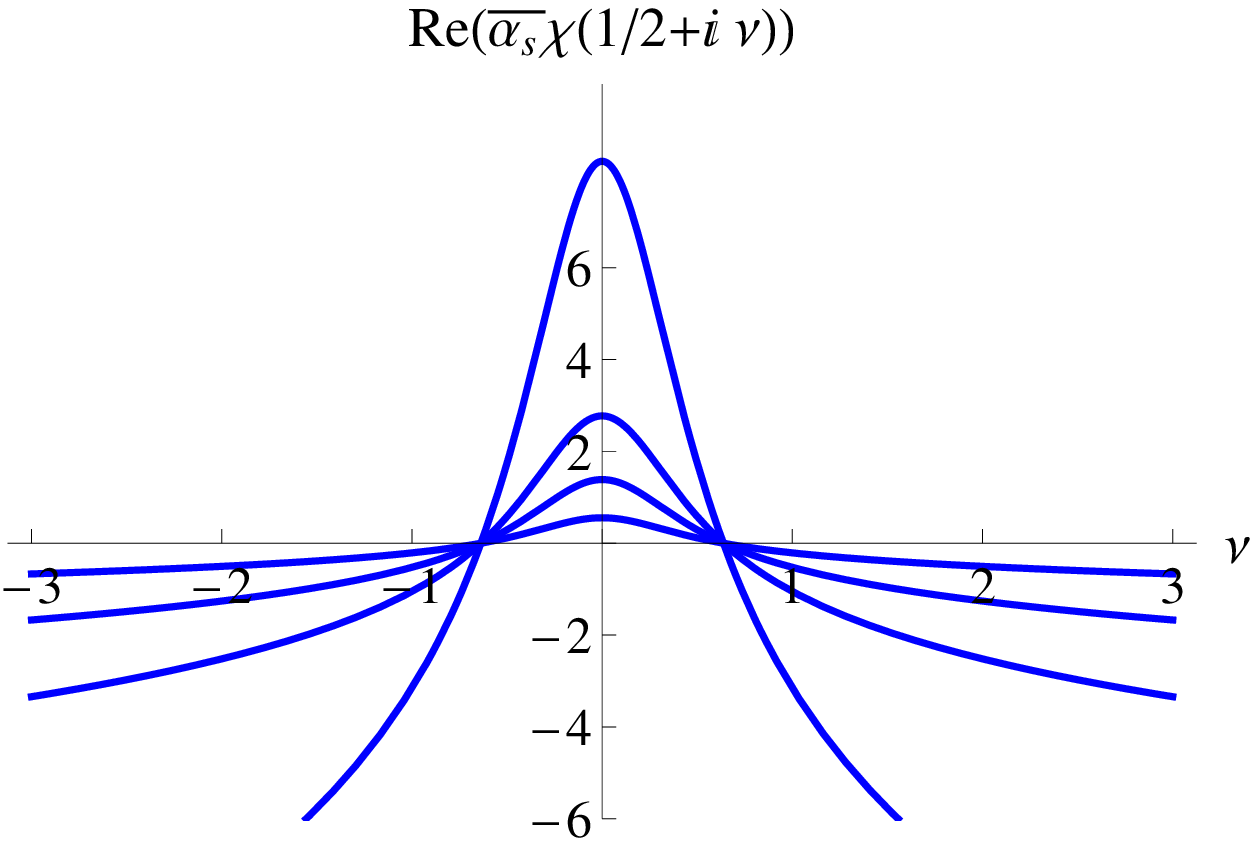}
    }

\end{picture}

\vspace{3cm}
\caption{\em \small Left: BFKL characteristic function multiplied by $\bar{\alpha}_s$ along the real axis. Right: BFKL characteristic function multiplied by $\bar{\alpha}_s$ along imaginary axis. The evaluation is for $\bar{\alpha}_s$=0.2, 0.5, 1, 3. Increasing values of the $y$ axis indicate the direction of the growth of $\bar{\alpha}_s$}.
\label{fig:plotBFKL}
\end{figure}

In this section, we review some basics of the integral form of the BFKL equation at LO and its solution in a diffusion approximation \cite{Forshaw:1997dc,Barone:2002cv}. The diffusive form of the BFKL equation, as it turns out later, is the one that describes the infinite strong coupling regime when kinematical constraint \cite{Kwiecinski:1996td,Andersson:1995jt} and DGLAP corrections are imposed.
The forward BFKL equation written for the unintegrated gluon density in the integral form reads
\be
f(x,k^2)=f_0(x,k^2)+\bar{\alpha}_sk^2\int_{x/x_0}^1\frac{dz}{z}\int_0^\infty\frac{dl^2}{l^2}\left[\frac{f(x/z,l^2)-f(x/z,k^2)}{|l^2-k^2|}+\frac{f(x/z,k^2)}{\sqrt{4l^4+k^4}}\right],
\label{eq:BFKL1}
\ee
where $x$ is a longitudinal momentum fraction carried by the gluon, $k$ is the modulus of its transversal momentum, and we use $k^2$ to indicate that there is no angular dependence. The normalization of the gluon is such that in the double logarithmic limit one has the relation $xg(x,Q^2)=\int_0^{Q^2}dk^2f(x,k^2)/k^2$. The LO BFKL equation due to conformal invariance can be solved by the Mellin transform. The Mellin transform with respect to $x$ and its inverse read
\be
{\overline f}(\omega,k^2)=\int_0^{1}dx x^{\omega-1}f(x,k^2),\,\,\,\,\,\,\,
f(x,k^2)=\frac{1}{2\pi i}\int_{c-i\infty}^{c+i\infty}d\omega x^{-\omega}{\overline f}(\omega,k^2).
\ee
Applying it to both sides of (\ref{eq:BFKL1}) and using \footnote{To simplify the notation, we kept the same letter for the Mellin transform with respect to $k^2$.}
\be
{\overline f}(\omega,k^2)=\frac{1}{2\pi i}\int_{c-i\infty}^{c+i\infty}(k^2)^\gamma {\overline f}(\omega,\gamma)d\gamma
\ee
one obtains
\be
{\overline f}(\omega,\gamma)={\overline f}_0(\omega,\gamma)+\frac{\bar{\alpha}_s}{\omega}\chi(\gamma){\overline f}(\omega,\gamma),
\ee
where
\be \label{integralintercept}
\chi(\gamma)=\int_0^\infty\frac{du}{u}\left[\frac{u^{\gamma}-1}{|u-1|}+\frac{1}{\sqrt{4u^2+1}}\right].
\ee
The integral (\ref{integralintercept}) after evaluation gives
\be
\chi(\gamma)=2\psi(1)-\psi(1-\gamma)-\psi(\gamma),
\ee
where $\psi$ is a digamma function. As a result the solution to Eq. (\ref{eq:BFKL1}) can be written as
\be
f(x,k^2)=\frac{1}{2\pi i}\int d\gamma (k^2)^{\gamma}\frac{1}{2\pi i}\int d\omega x^{-\omega}\frac{\omega {\overline f}_0(\omega,\gamma)}{\omega-\overline{\alpha_s}\chi(\gamma)}.
\ee
Taking for the boundary condition
\be
f_0(x,k^2)=f(x_0,k^2),
\ee
which, in the Mellin transform, corresponds to
\be
{\overline f}_0(\omega,\gamma)=\frac{f(x_0,\gamma)}{\omega},
\ee
we arrive at the following expression:
\be
f(x,k^2)=\frac{1}{2\pi i}\int d\gamma (k^2)^{\gamma}f(x_0,\gamma)\left(\frac{x}{x_0}\right)^{-\bar{\alpha}_s\chi(\gamma)}.
\ee
In order to evaluate the integral above one needs to know the characteristic function along the imaginary axis in the $\gamma$ plane. The characteristic function is an analytic function given by the formula above, and its value along the imaginary axis can be easily obtained.  We present the characteristic function along the real and imaginary axis for various values of the coupling constant in Fig. (\ref{fig:plotBFKL}). One sees that at large values of the coupling constant the characteristic function diverges, and from this one concludes that the LO BFKL equation cannot be naively extended to the large coupling constant regime.
Knowing that the characteristic function has a saddle point along the $\gamma= 1/2+i\nu$ contour, we can write the solution as
\be
f(x,k^2)=\frac{1}{2\pi}\int_{-\infty}^{\infty} d\nu (k^2)^{1/2+i\nu}{\overline f}(x_0,1/2+i\nu) x^{-\bar{\alpha}_s \chi(1/2+i\nu)}.
\ee
The dimensionful unintegrated gluon density reads
\be
{\cal F}(x,k^2)={\cal F}(x_0,1/2)\frac{1}{\sqrt{4\pi\ln(x_0/x)1/2\lambda^{\prime}}}e^{\lambda\ln(x_0/x)-1/2\ln(k^2/k_0^2)}e^{\frac{-\ln(k^2/k_0^2) ^2}{4\,1/2\lambda^{\prime}\ln(x_0/x)}},
\ee
where ${\cal F}(x_0,1/2)=f(x_0,1/2)/k^2$ and $\chi(1/2+i\nu)\approx\lambda-\frac{1}{2} \lambda^{\prime}\nu^2$ with $\lambda=\bar{\alpha}_s4\ln2$ and $\lambda^{\prime}=\bar{\alpha}_s\zeta(3)$.
From this explicit form, one may extract the coefficients of the diffusion equation \footnote{For the details about the diffusion equation we refer the reader to Appendix \ref{appdiffeq}.}
\be \label{diffusioneq}
\partial_Y{\cal F}(Y,\rho)=\frac{1}{2} \lambda^{\prime} \partial^2_\rho {\cal F}(Y,\rho)+\frac{1}{2}\lambda^{\prime}\partial_\rho{\cal F}(Y,\rho)+(\lambda+\lambda^{\prime}/8){\cal F}(Y,\rho).
\ee

In the above expression we used the following variables:
\be\label{igrek}
Y=\ln \frac{x_0}{x},
\ee
\be \label{ro}
\rho = \ln \frac{k^2}{k_0 ^2}.
\ee
The variables (\ref{igrek}) and (\ref{ro}) are convenient, because they allow one to write Eq. (\ref{diffusioneq}) in a simple diffusive form.

\section{The BFKL equation with higher order corrections and the gluon density in the whole range of coupling constant}\label{BFKLstrong}
\begin{figure}[t!]
  \begin{picture}(30,30)
    \put(30, -80){
      \includegraphics{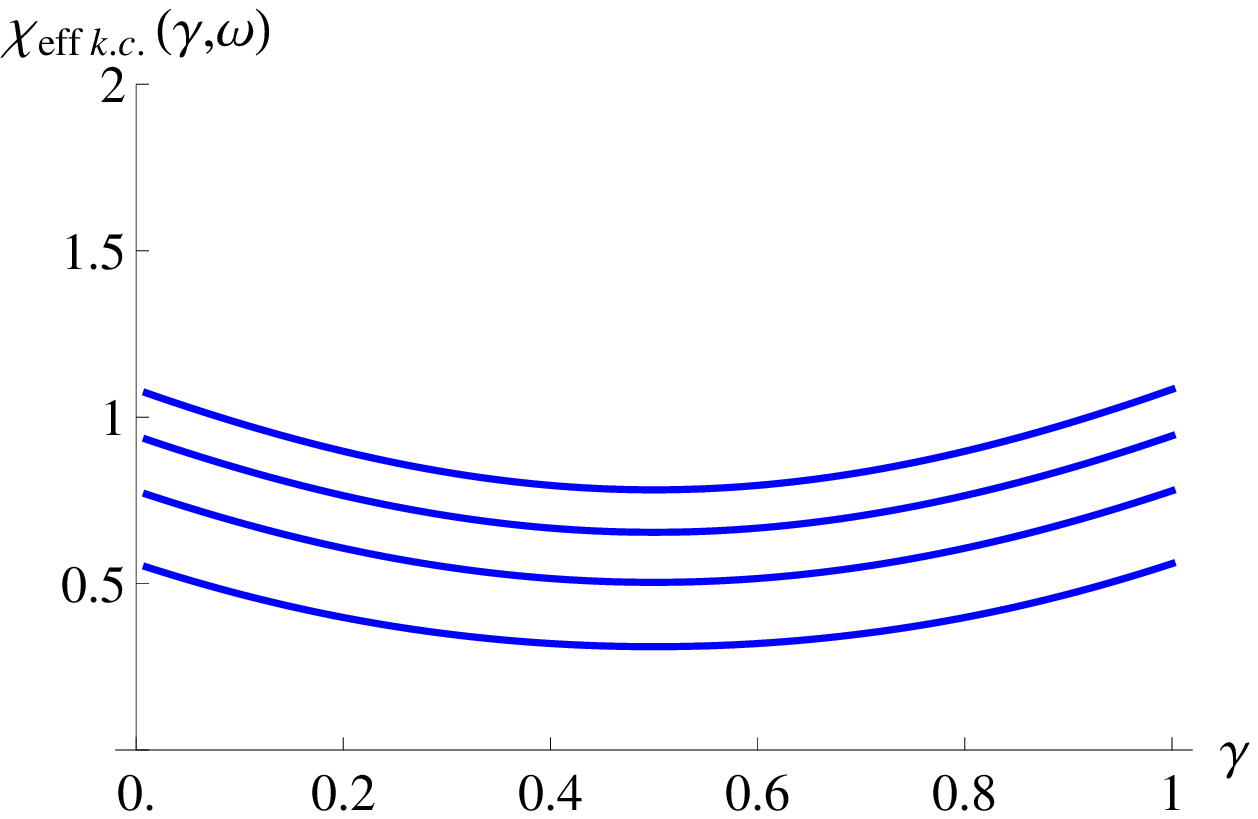}
    }

 \put(280, -80){
      \includegraphics{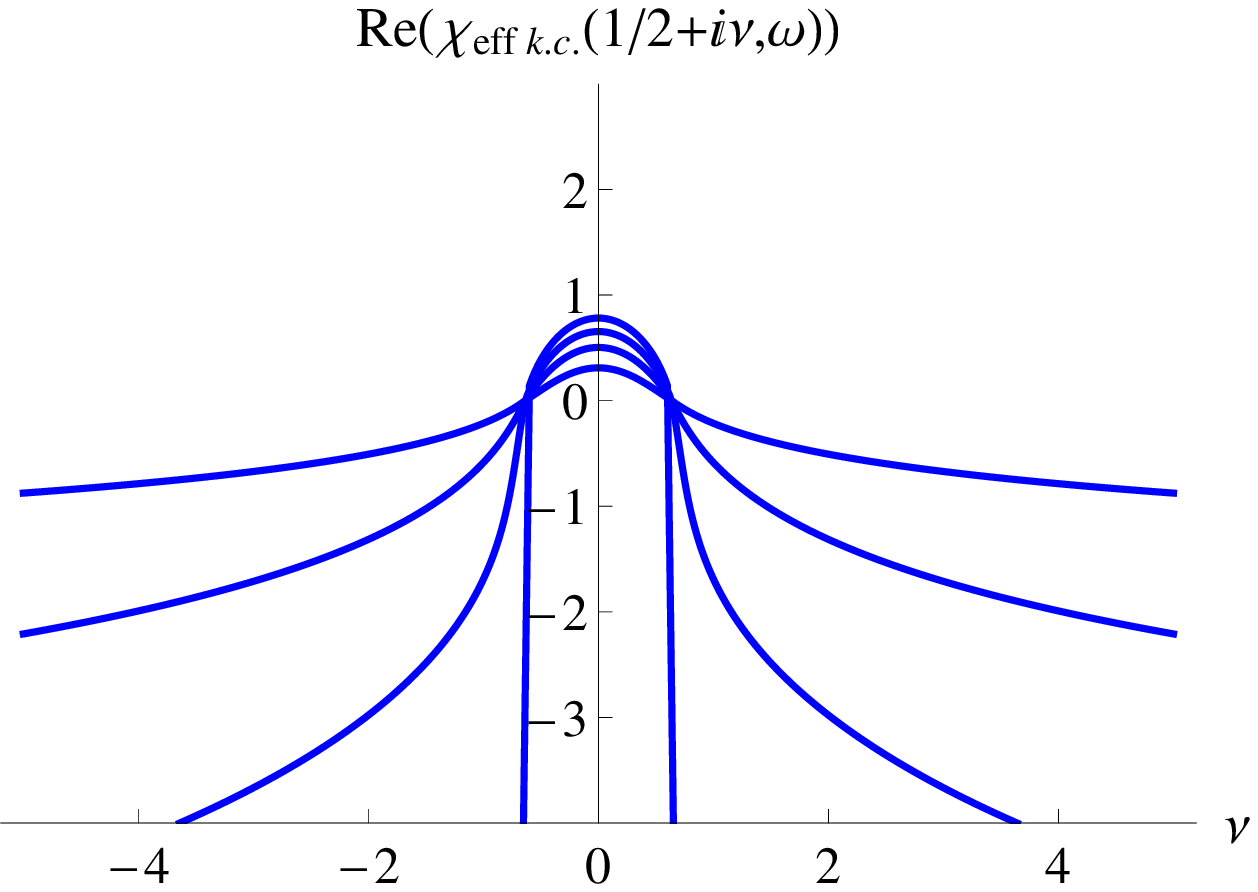}
    }

\put(30, -280){
      \includegraphics{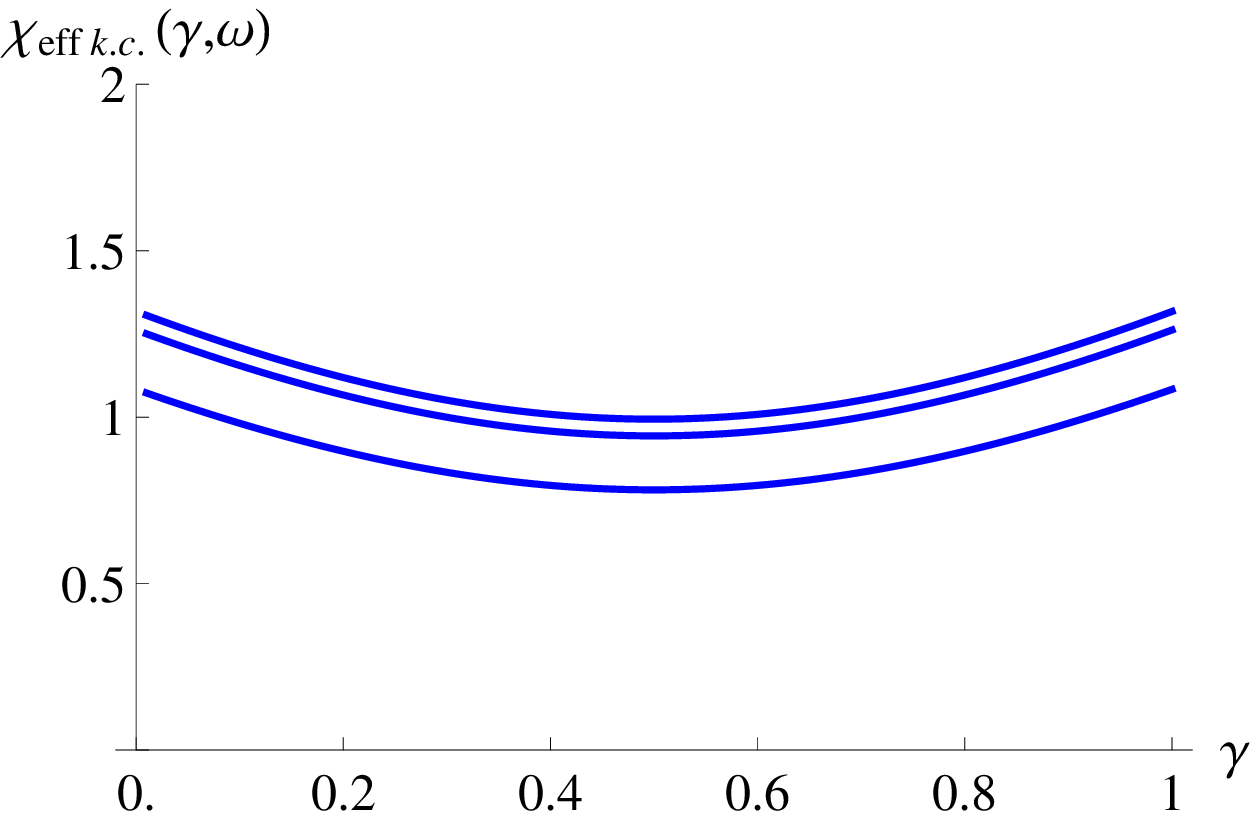}
    }

\put(280, -280){
      \includegraphics{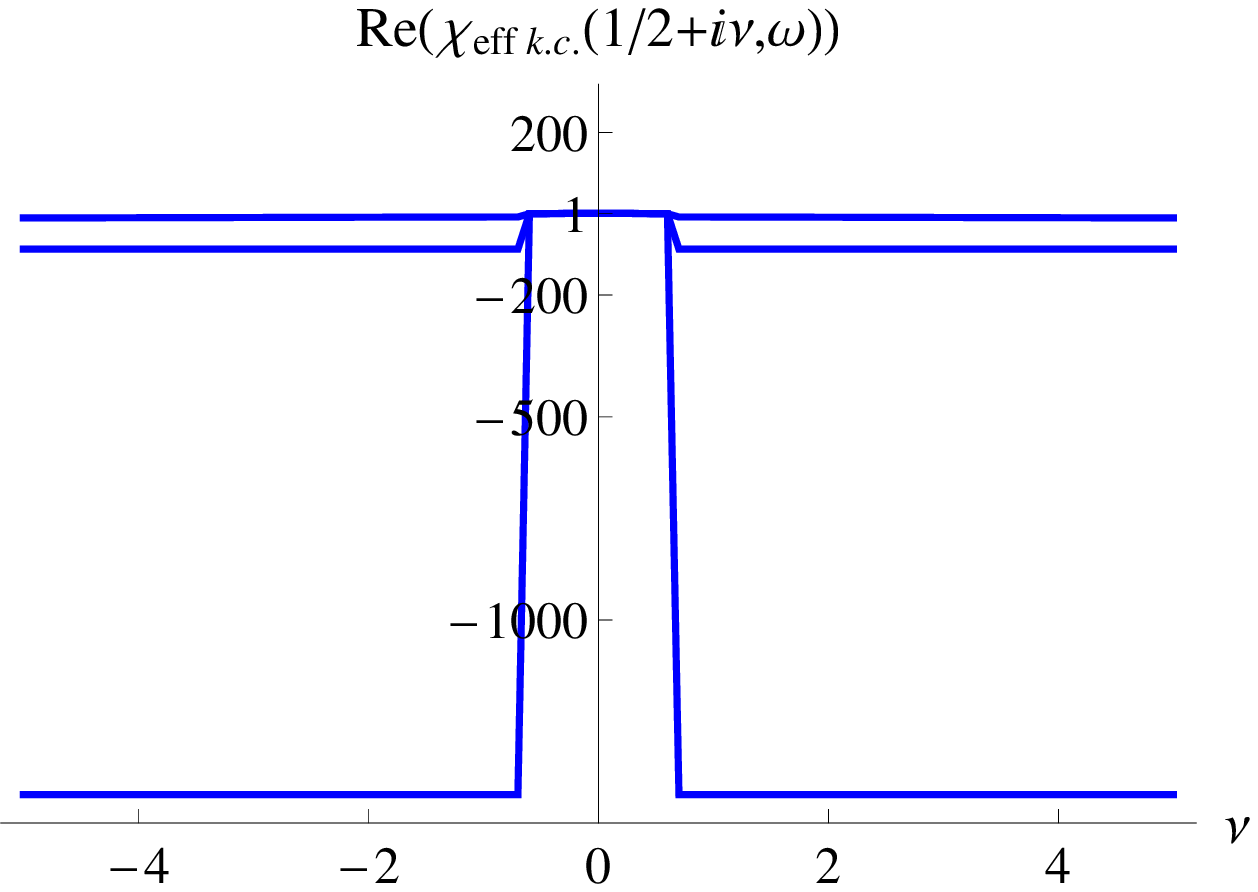}
    }

\end{picture}
\vspace{10cm}
\caption{\em \small Kinematical constraint effects. Upper right plot: Function $\chi_{eff\, k.c.}(\gamma,\omega)$ along the real contour for $\bar{\alpha}_s=0.2,0.5,1,2$.
 Upper left plot: Function $\chi_{eff\, k.c.}$ along the imaginary contour for $\bar{\alpha}_s=0.2,0.5,1,2$. Lower left plot:  Function $\chi_{eff\, k.c.}(\gamma,\omega)$ along the real contour for $\bar{\alpha}_s=2,10,100$.  Lower right plot:  Function $\chi_{eff\, k.c.}(\gamma,\omega)$ along the real contour for $\bar{\alpha}_s=2,10,100$.}
\label{fig:plotvel2}
\end{figure}

\begin{figure}[t!]
  \begin{picture}(30,30)
    \put(20, -80){
      \includegraphics{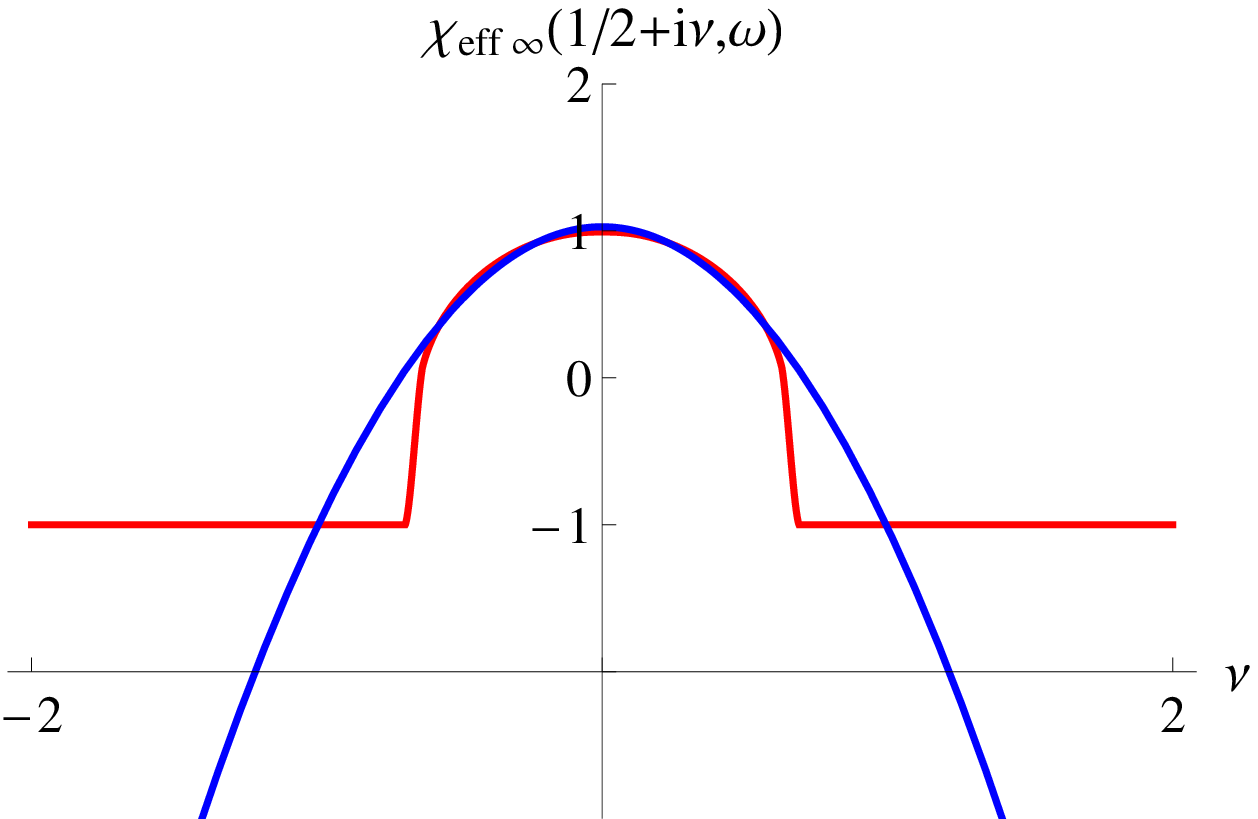}
    }

\put(280, -80){
      \includegraphics{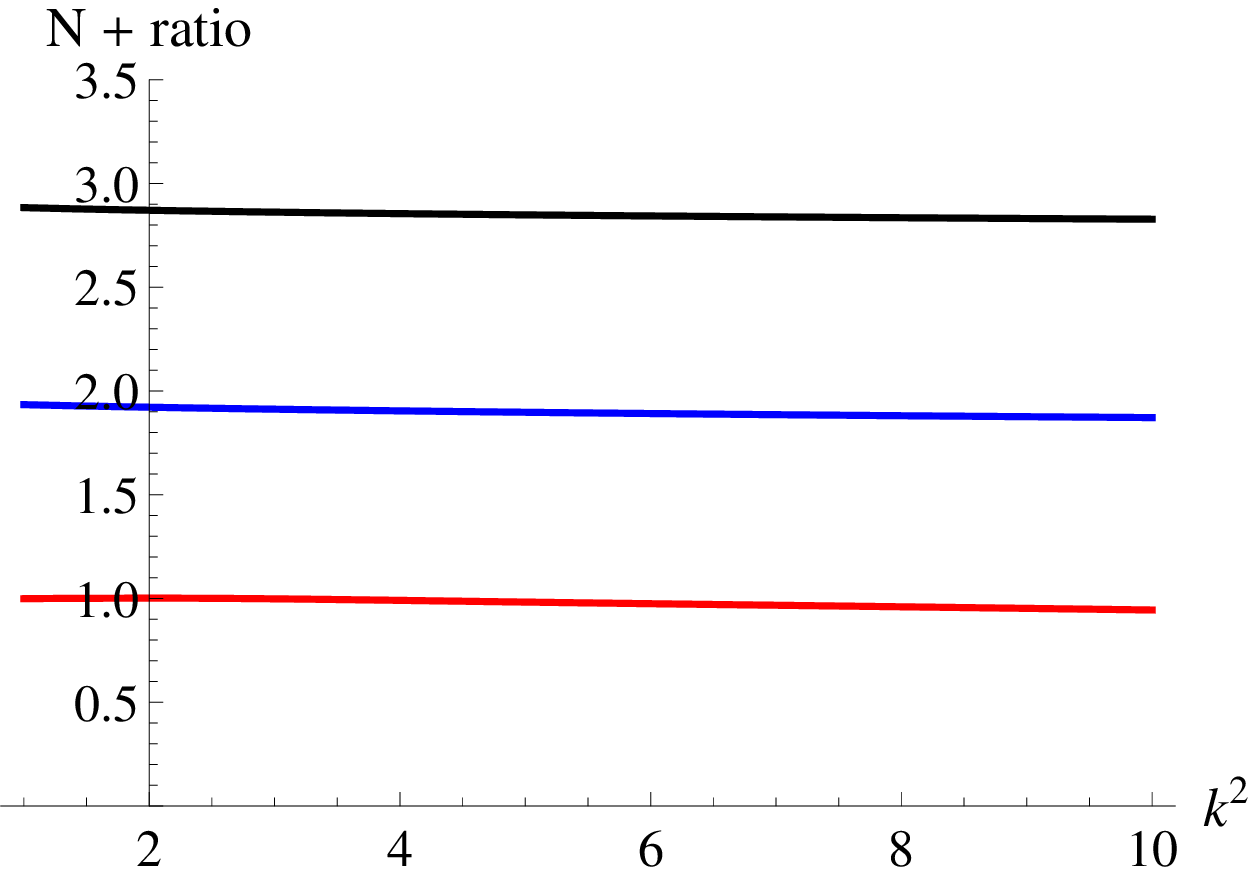}
    }

\end{picture}

\vspace{3cm}
\caption{\em \small Left: Comparison of $\omega$ from formulas (\ref{eq:fit10}) and (\ref{eq:fitquad}). Right: Ratio of gluon densities for different values of $x$ as obtained from  (\ref{eq:fit10}) and (\ref{eq:fitquad}. In order to visualize the tiny effect of dependence on the exact integration path, we plot ratios and shift the appropriate curve by 2 and by 3, respectively.}
\label{fig:plotvel3}
\end{figure}

The BFKL equation has been obtained at NLO accuracy in Refs. \cite{Fadin:1998py,Balitsky:2008zza} and recently solved in Ref. \cite{Chirilli:2013kca}. However, it turns out that, in order to make the eigenvalue of the kernel stable, one needs to perform resummations of corrections to infinite order \cite{Salam:1999cn}. One source of such corrections is provided by the so-called kinematical constraint effects. The LO BFKL equation has been derived by assuming strong ordering in energies of gluons emitted in the $s$ channel. However, the integral over the transversal momentum is in principle unconstrained. The way to improve the situation is to demand that the emitted gluons, when the longitudinal momentum fraction
$z\rightarrow 1$, are on shell. This puts certain restrictions on the transversal momentum.
The kinematical constraint refined equation reads
\begin{align}
f(x,k^2)&=f_0(x,k^2)\nonumber \\
        &+\bar{\alpha}_sk^2\int_{x}^1\frac{dz}{z}\int_0^\infty\frac{dl^2}{l^2}\left[\frac{f(x/z,l^2)\theta(l-kz)\theta(k/z-l)-f(x/z,k^2)}{|l^2-k^2|}+\frac{f(x/z,k^2)}{\sqrt{4l^4+k^4}}\right],
\label{eq:bfklint}
\end{align}
Since it does not break the conformal invariance, the improved equation can be again solved by the Mellin transform technique. Performing the transform with respect to $x$, we get\footnote{For the technical details about the evaluation of such integrals, we refer the reader to Appendix \ref{appint}. }
\be
{\overline f}(\omega,k^2)={\overline f}_0(\omega,k^2)+\bar{\alpha}_s\frac{k^2}{\omega}\int_0^\infty\frac{dl^2}{l^2}\left[\frac{{\overline f}(\omega,l^2)\theta(l-kz)\theta(k/z-l)-{\overline f}(\omega,k^2)}{|l^2-k^2|}+\frac{{\overline f}(\omega,k^2)}{\sqrt{4l^4+k^4}}\right].
\ee
We have to combine contributions coming from both $l^2>k^2$ and $l^2<k^2$. Taking that into account we perform the Mellin transform with respect to $k^2$ and obtain
\be
{\overline f}(\omega,\gamma)={\overline f}_{0}(\omega,\gamma)+\frac{\bar{\alpha}_s}{\omega}\chi(\gamma,\omega){\overline f}(\omega,\gamma),
\label{eq:solutionBFKL}
\ee
where
\be
\chi_{k.c.}(\gamma,\omega)=\int_0^{\infty}\frac{du}{u}\left[\frac{u^{\gamma+\omega/2}\theta(1-u)+u^{\gamma-\omega/2}\theta(u-1)-1}{|1-u|}+\frac{1}{\sqrt{4u^2+1}}\right].
\ee
We relegate the details of evaluation of the integrals to Appendix \ref{appint}; the final result is
\be
\chi_{k.c.}(\gamma,\omega)=2\psi(1)-\psi(1-\gamma+\omega/2)-\psi(\gamma+\omega/2).
\ee
After rearranging (\ref{eq:solutionBFKL}) and using the inverse Mellin transforms with respect to $\omega$ and $\gamma$, we obtain
\be
f(x,k^2)=\frac{1}{2\pi i}\int d\gamma (k^2)^{\gamma}\frac{1}{2\pi i}\int d\omega x^{-\omega}\frac{\omega {\overline f}_0(\omega,\gamma)}{\omega-{\bar\alpha}_s\chi_{k.c.}(\gamma,\omega,)}.
\label{eq:lastsolBFKL}
\ee
Equation (\ref{eq:lastsolBFKL}) defines a transcendental equation, and its solution gives a modified energy dependence of the BFKL gluon density:
\be
\omega={\bar\alpha}_s\chi_{k.c.}(\gamma,\omega)\equiv\chi_{eff\,k.c.}(\gamma,\omega).
\label{eq:kinconstraint}
\ee
Solving Eq. (\ref{eq:kinconstraint}), we see that the kinematical effects limit the growth of the eigenvalue to large values. We notice, however, that the eigenvalue along the imaginary axis is unlimited from below (see Fig. \ref{fig:plotvel2}), where we solved (\ref{eq:kinconstraint}) along the imaginary axis, i.e.,
\be
\omega=\Re\left(\chi_{eff\,k.c.}(1/2+i\nu,\omega)\right)
\ee
\begin{figure}[t!]
  \begin{picture}(30,30)
    \put(30, -80){
      \includegraphics{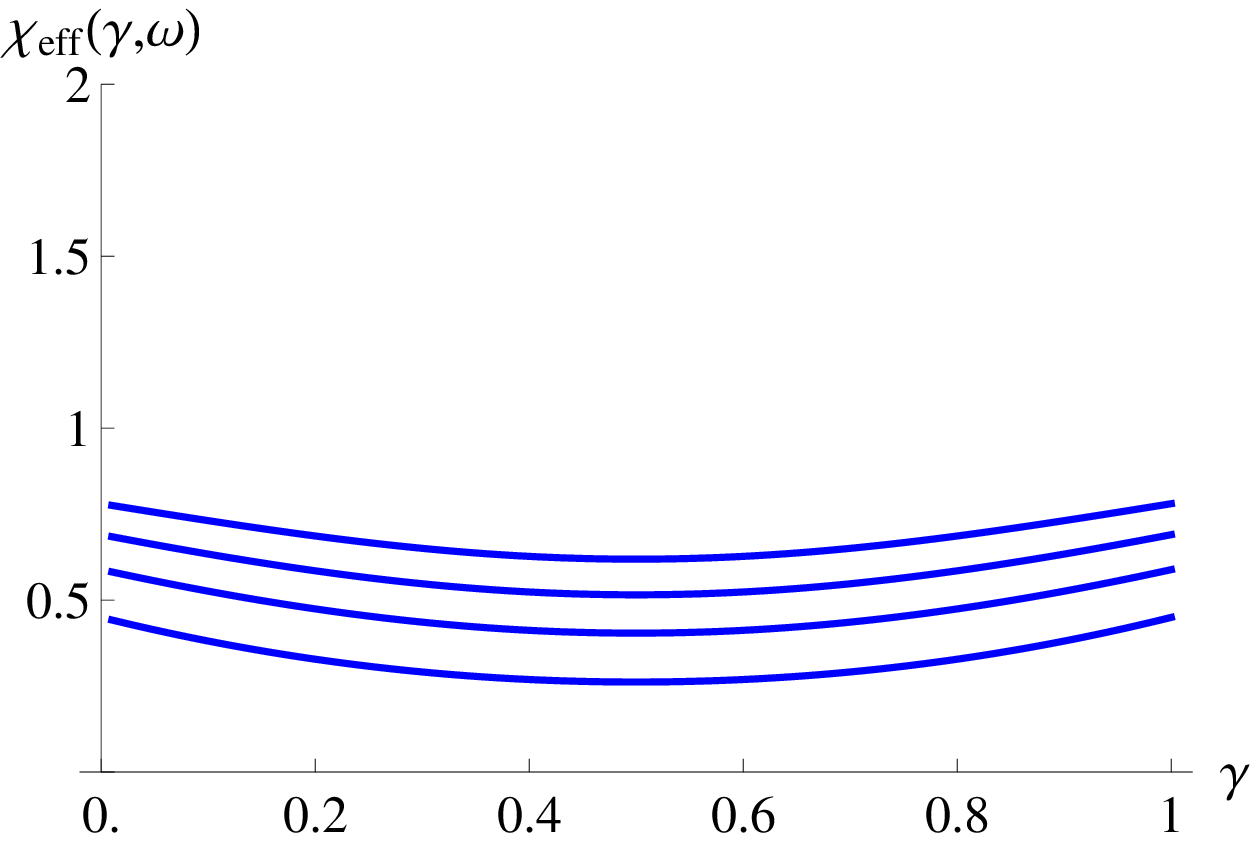}
    }

 \put(280, -80){
      \includegraphics{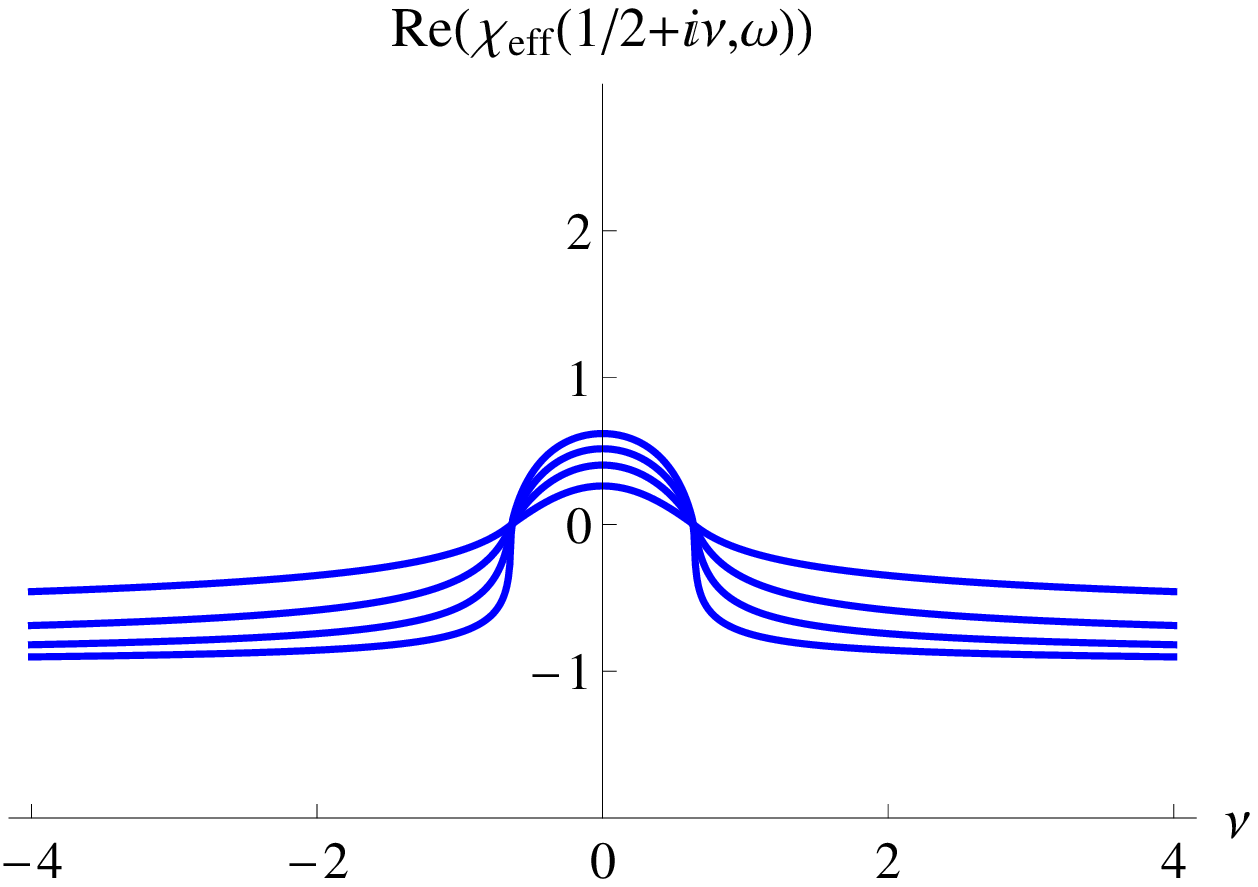}
    }

\put(30, -280){
      \includegraphics{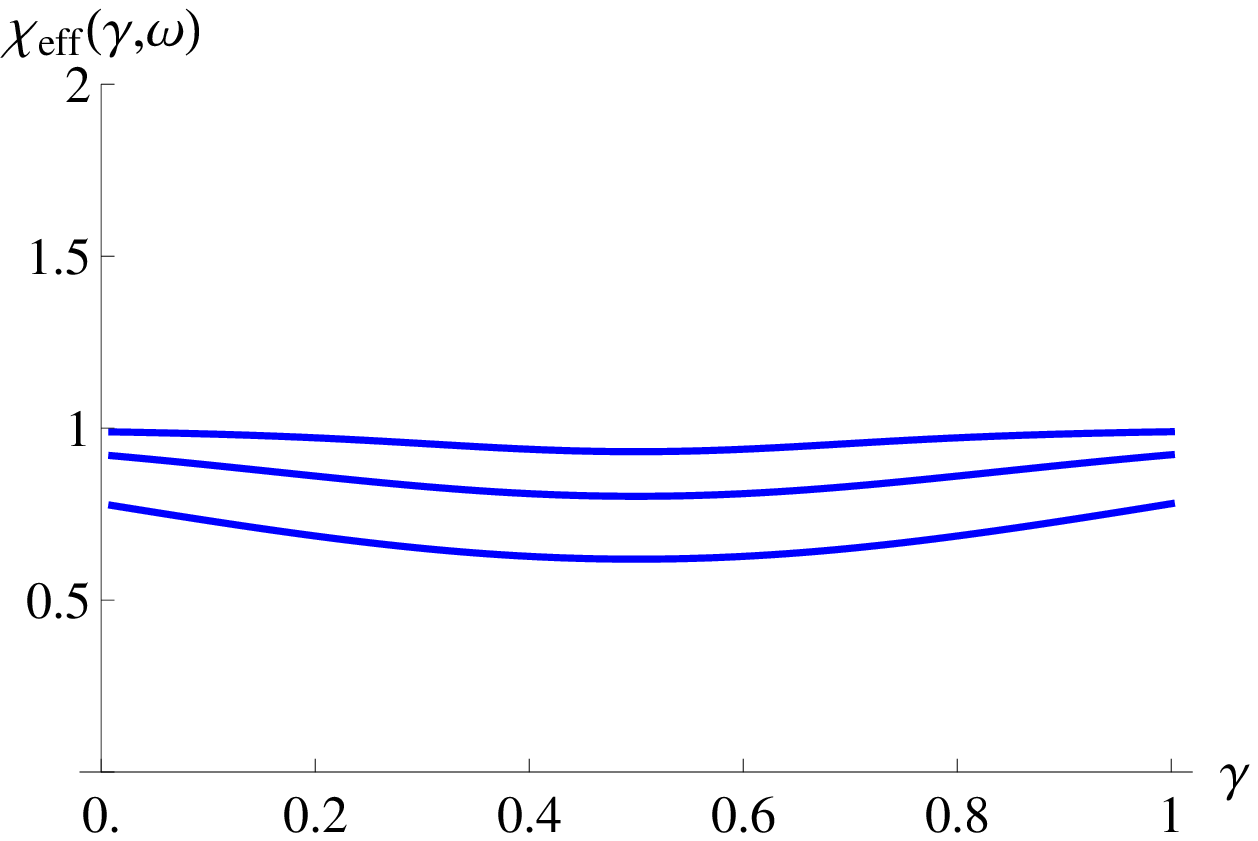}
    }

\put(280, -280){
      \includegraphics{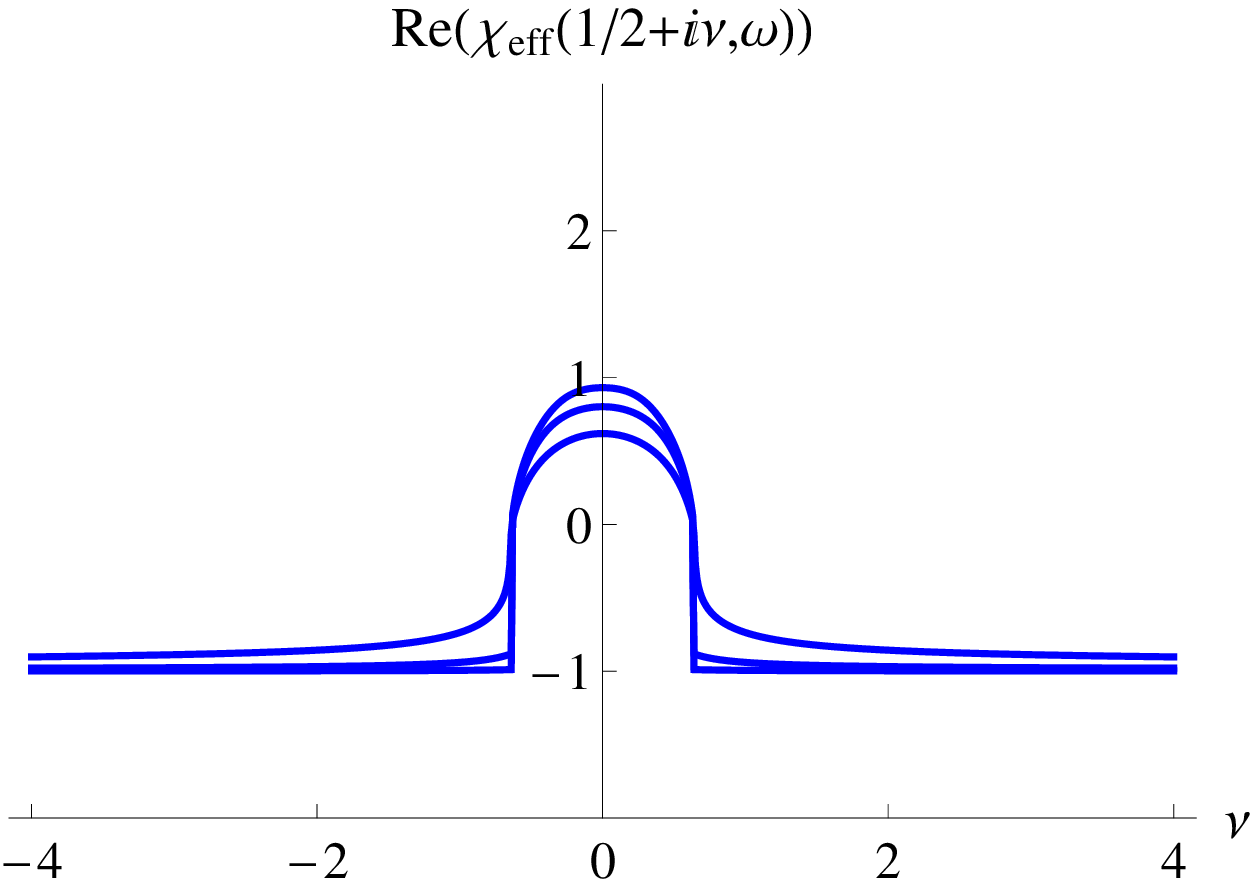}
    }

\end{picture}
\vspace{10cm}
\caption{\em \small  Kinematical constraint effects and resummation effects. Upper right plot: Function $\chi_{eff}(\gamma,\omega)$ along the real contour for $\bar{\alpha}_s=0.2,0.5,1,2$.
 Upper left plot: Function $\chi_{eff}(\gamma,\omega)$ along the imaginary contour for $\bar{\alpha}_s=0.2,0.5,1,2$. Lower left plot:  Function $\chi_{eff}(\gamma,\omega)$ along the real contour for $\bar{\alpha}_s=2,10,100$.  Lower right plot:  Function $\chi_{eff}(\gamma,\omega)$ along the real contour for $\bar{\alpha}_s=2,10,100$.}
\vspace{1.5cm}
\label{fig:kinres}
\end{figure}
It has been suggested in Ref. \cite{Stasto:2007uv} that, in order to have a more complete treatment of the contribution of higher orders from the point of view of BFKL, one modifies Eq. (\ref{eq:kinconstraint}) to the following one:
\be
\frac{1}{{\bar \alpha}_s}=\gamma^{(0)}(\omega)\chi_{k.c.}(\gamma,\omega),
\label{eq:effeigen}
\ee
where
\be
\gamma^{(0)}(\omega)=\frac{1}{\omega}+A(\omega)
\ee
is the LO DGLAP anomalous dimension and
\be
A(\omega)=-\frac{1}{\omega+1}+\frac{1}{\omega+2}-\frac{1}{\omega+3}-\psi(2+\omega)+\psi(1)+\frac{11}{12}.
\ee
\begin{figure}[t!]
  \begin{picture}(30,30)
    \put(30, -80){
      \includegraphics{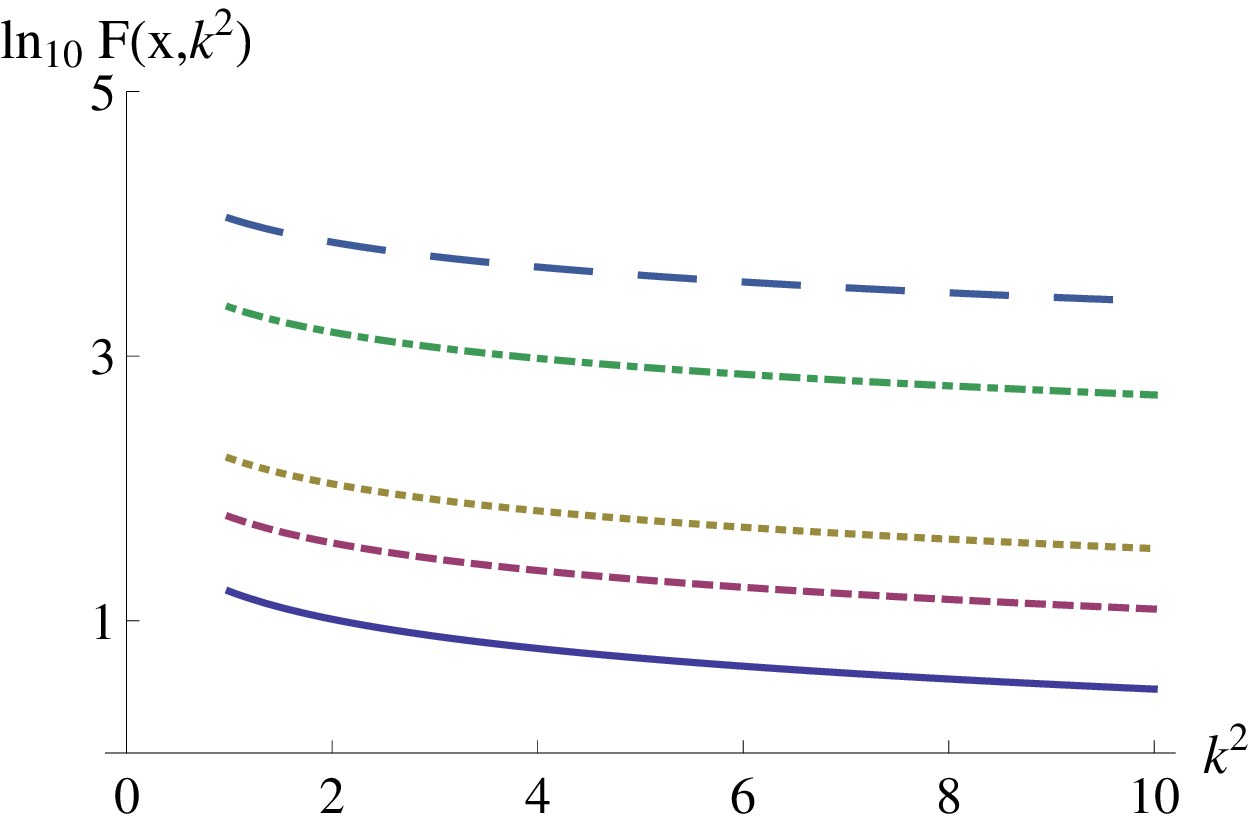}
    }

 \put(280, -80){
      \includegraphics{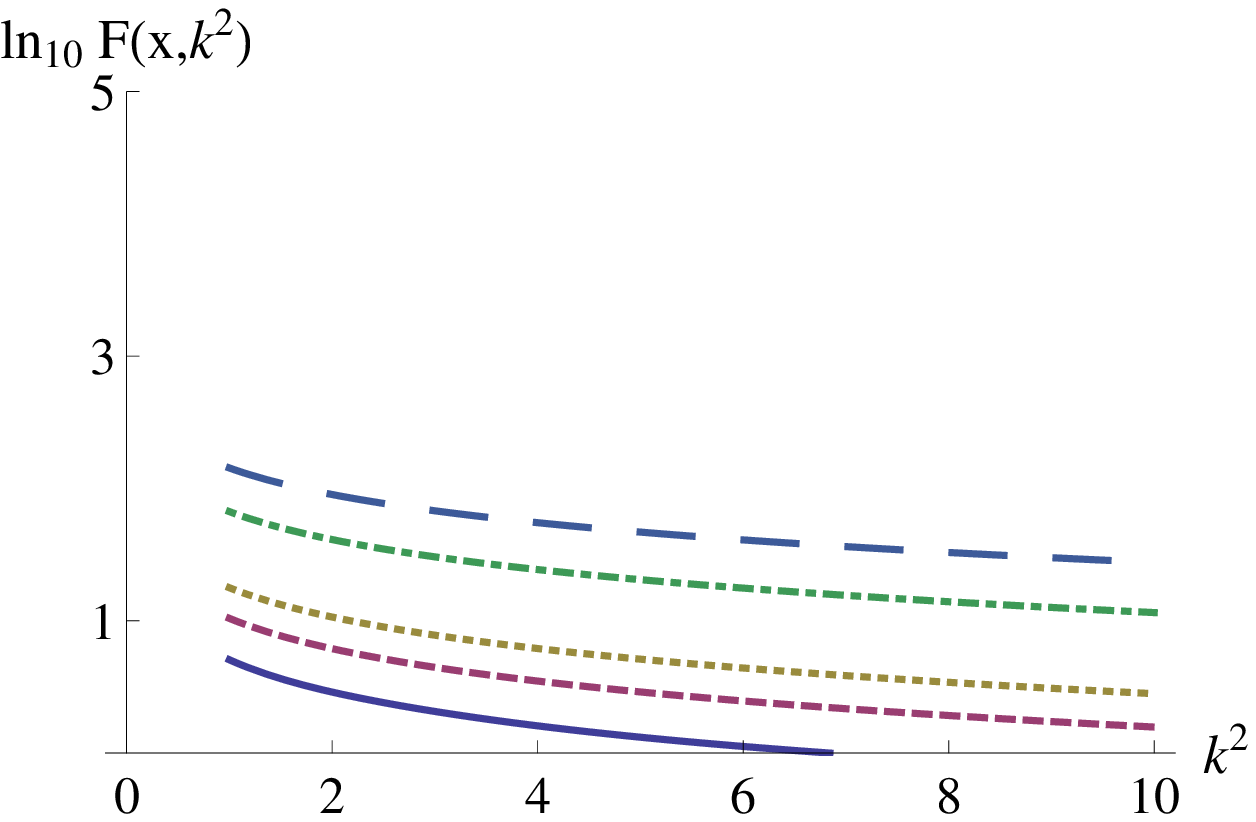}
    }

\end{picture}
\vspace{3cm}
\caption{\em \small  Gluon density obtained for various values of the coupling constant ${\bar\alpha}_s\!=\!0.2,\,0.5,\,1,\,10, \,10^4$. The densities decrease for a smaller coupling constant. Left, $x=10^{-6}$; right, $x=10^{-4}$. }
\vspace{1cm}
\label{fig:plotvel5}
\end{figure}
Equation (\ref{eq:effeigen}) provides a resummation of DGLAP gluon anomalous dimension at LO (missing in BFKL) and kinematical effects. It can be written as an effective eigenvalue equation in the form
\be
\omega= \chi_{eff}(\gamma,\omega),
\ee
with
\be
\chi_{eff}(\gamma,\omega)=\bar{\alpha}_s\chi_{k.c.}(\gamma,\omega)\left(1+A\omega\right).
\ee
The crucial behavior, providing the gravitonlike intercept, is the vanishing of the eigenvalue function when $\omega\rightarrow 1$.
As a practical application of the result obtained in Ref. \cite{Stasto:2007uv}, we ask a question about the  properties of the gluon density while evaluated at the increasing values of strong coupling.
To obtain the gluon density we know from previous sections that we need to know the eigenvalue function along the imaginary axis to perform the inverse Mellin transform.
We see (Fig. \ref{fig:kinres}) that the additional contributions stabilize completely the eigenvalue and allow for the investigations of the BFKL in the whole spectrum of the coupling constant. This stems from the fact that  there is no divergency after taking the limit $\bar{\alpha}_S\rightarrow\infty$.
Keeping this in mind, we can ask, what is the shape of the gluon density as we flow to the larger values of the coupling constant? Naively, one can think that if the coupling constant increases, the number of gluons vanishes or at least it is constant. Below, we show that this is not the case, the gluon density grows, and we get infinitely many soft gluons. Thus, in order to achieve the stabilization, one has to include some additional effects, most probably of the non-linear type.
In order to obtain the gluon density, we interpolate the solution and integrate it numerically. The resulting gluon density for different values of the coupling constant is shown in Fig. \ref{fig:plotvel5}.
A simple structure of the solution of the eigenfunction equation for $\omega$ can be parameterized in a polynomial of $\nu$ where $A_n$ are fit parameters:
\be
\chi_{eff\,\infty}(\omega,1/2+i\nu)=\sum _{n=-M}^{N}A_n\nu^n.
\ee
Applying this prescription, we obtain at the infinite value of the strong coupling \footnote{It works also at finite values.} the following formula:
\be
\chi_{eff\,\infty}(\omega,1/2+i\nu)= P_{10} (\nu)
   \theta (\nu +0.683) \theta (0.683\, -\nu )-\theta (-\nu -0.683)-\theta (\nu -0.683),
\label{eq:fit10}
\ee
where the tenth-order polynomial $P_{10} (\nu)$ takes the form
\be
P_{10} (\nu)=0.998873-2.01319\nu ^2 +15.9008\nu ^4-154.039 \nu ^6+540.208 \nu ^8-657.203 \nu ^{10}.
\ee
The form of the eigenvalue function used in order to evaluate the gluon density can be simplified. By inspection we see (Fig. \ref{fig:plotvel3}, right) that if for the evaluation of the gluon density we use the simplified fit
\be
\chi_{eff\,\infty}(\omega,1/2+i\nu)=1.02795\, -2.04635 \nu ^2\equiv\lambda_{st}-\frac{1}{2}\lambda_{st}^{\prime}\nu^2,
\label{eq:fitquad}
\ee
the resulting gluon almost does not change. The reason for this is that the net contribution to the integral from regions of the eigenvalue function where it is negative and where the functions start to differ is negligible (see Fig. \ref{fig:plotvel3}, left).
Formula (\ref{eq:fitquad}) can be used to obtain analytically a solution of the BFKL equation in the strong coupling regime and to deduce a partial differential equation which it obeys:
\be
\partial_Y\Phi(Y,\rho)=\frac{1}{2}\lambda_{st}^{\prime} \partial^2_\rho \Phi(Y,\rho)+\frac{1}{2}\lambda_{st}^{\prime}\partial_\rho\Phi(Y,\rho)+(\lambda_{st}+\lambda_{st}^{\prime}/8)\Phi(Y,\rho),
\label{eq:BFKLstrong}
\ee
where the values are read off from formula (\ref{eq:fitquad}) $\lambda_{st}^{\prime}=4.08, \,\lambda_{st}=1.02$

\section{AdS/CFT and the Pomeron}\label{AdSpomeron}
The main message that comes from our discussion so far is that the diffusive behavior is not particular to the weak coupling regime of the BFKL evolution. In fact it is the dominant contribution in the strong coupling as well. This is not an artifact of the model we are considering. A similar observation was done in the context of gauge and gravity duality\cite{Maldacena:1997re,Witten:1998qj,Gubser:1998bc}, where an analytic expression for the gluon distribution related function (or an evolution kernel) was derived in $\mathcal{N}=4$ SYM theory:
\be
f_{PSBC}(\rho,s)\approx \frac{1}{\sqrt{4\pi \mathcal{D}Y}}e^{j_0 Y} e ^\frac{-\rho^2}{4\mathcal{D}Y},
\ee
where
\be
j_0=2-{2\over\sqrt\lambda}
+ O(1/\lambda)\
, \quad {\cal D}= {1\over 2\sqrt\lambda} + O(1/\lambda)\ \ .
\ee
and we introduced the notation $f_{PSBC}$ to indicate that the function is not directly the gluon distribution function but related to it after rescaling by $k^{1/2}$ \cite{Askew:1993jk}:
\be
\frac{\partial f_{PSCB}(Y,\rho)}{\partial Y}=\mathcal{D} \frac{\partial ^2 f_{PSCB}(Y,\rho)}{\partial \rho ^2}+j_0 f_{PSCB}(Y,\rho).
\label{eq:diffads}
\ee
This is a very important result, since it gives an independent justification for Eq. (\ref{eq:fitquad}). Even though the BFKL equation seems to have a regular limit at large coupling, \textit{a priori}, it is not obvious that the structure of the evolution is the same at the actual strong coupling limit. Given the behavior (\ref{eq:diffads}), we expect that the diffusive evolution at large values of the coupling is a very universal phenomenon shared by various gauge theories. Therefore, we conclude that the resummation procedure presented before is consistent and allows one to investigate various questions related to the evolution of parton densities. Moreover, with the insight coming from gauge and gravity correspondence it might be even possible to extend the evolution equation to the whole spectrum of the coupling, at least for $\mathcal{N}=4$ SYM. We note that in $\mathcal{N}=4$ SYM has several remarkable properties, such as integrability of the evolution equation or the so-called maximal transcendentality property \cite{Beisert:2010jr}. These properties allow one to calculate the spectrum of anomalous dimensions and the Pomeron intercept to a high order in the inverse coupling expansion \cite{Janik:1999zk,Brower:2006ea,Cornalba:2007fs,Costa:2012cb,Kotikov:2013xu}. The full knowledge of anomalous dimensions is a necessary ingredient in the construction of resummed models valid in the whole spectrum of the coupling constant. We leave this problem for future research.

The most activity in the subject was devoted to the understanding of the linear BFKL equation at the large strong coupling limit, i.e., constructing the Pomeron intercept or the BFKL eigenfunction. At present, the detailed studies of unitarization are beyond reach within string theory, as one has to resum multiloop string amplitudes to all orders. However, under some simplified assumptions, the saturation line has been extracted from holographic considerations
\cite{Hatta:2007he,Mueller:2008bt,Kovchegov:2010uk,Cornalba:2008sp,Cornalba:2010vk}.
Our approach developed in the previous sections allows one to construct directly the gluon density function at strong coupling, the object which is ultimately used in calculating properties of various final states, for instance, momentum and rapidity spectra of jets or hadrons.
Furthermore, since the BFKL evolution equation is linear and diffusive it has a natural non-linear extension that governs saturation physics. We expect that loop corrections to string amplitudes can be resummed via some non-linear diffusion equation. We propose to investigate the BFKL nonlinear extensions in the limit of large strong coupling which we believe possess some universal features.
Indeed, in the next section, we will show how saturation physics can be included and that we get an agreement with the holographic result. This is a striking feature of the strongly coupled physics: Despite the differences between QCD and $\mathcal{N}=4$ SYM, they seem to share common properties such as validity of diffusive approximation and quantitatively similar saturation properties.

\section{The BK equation in the limit of infinite coupling constant $\bar{\alpha} _s$}\label{nonlineraeq}
The linear BFKL evolution equation misses a very important aspect of the high-energy scattering, namely, the saturation physics.  As pointed out in the introduction, several approaches were constructed in order to include non-linear effects, like multiple scattering and gluon saturation, responsible for the unitarization of the scattering amplitudes.  A particularly useful and simple enough approach to unitarize the cross section is the Balitsky-Kovchegov (BK) equation,
which reads:
\be
\Phi(x,k^2)=\Phi_0(x,k^2)+\Phi_1(x,k^2),
\label{eq:BFKL}
\ee
where
\begin{figure}[t!]
  \begin{picture}(30,30)
    \put(30, -80){
      \includegraphics{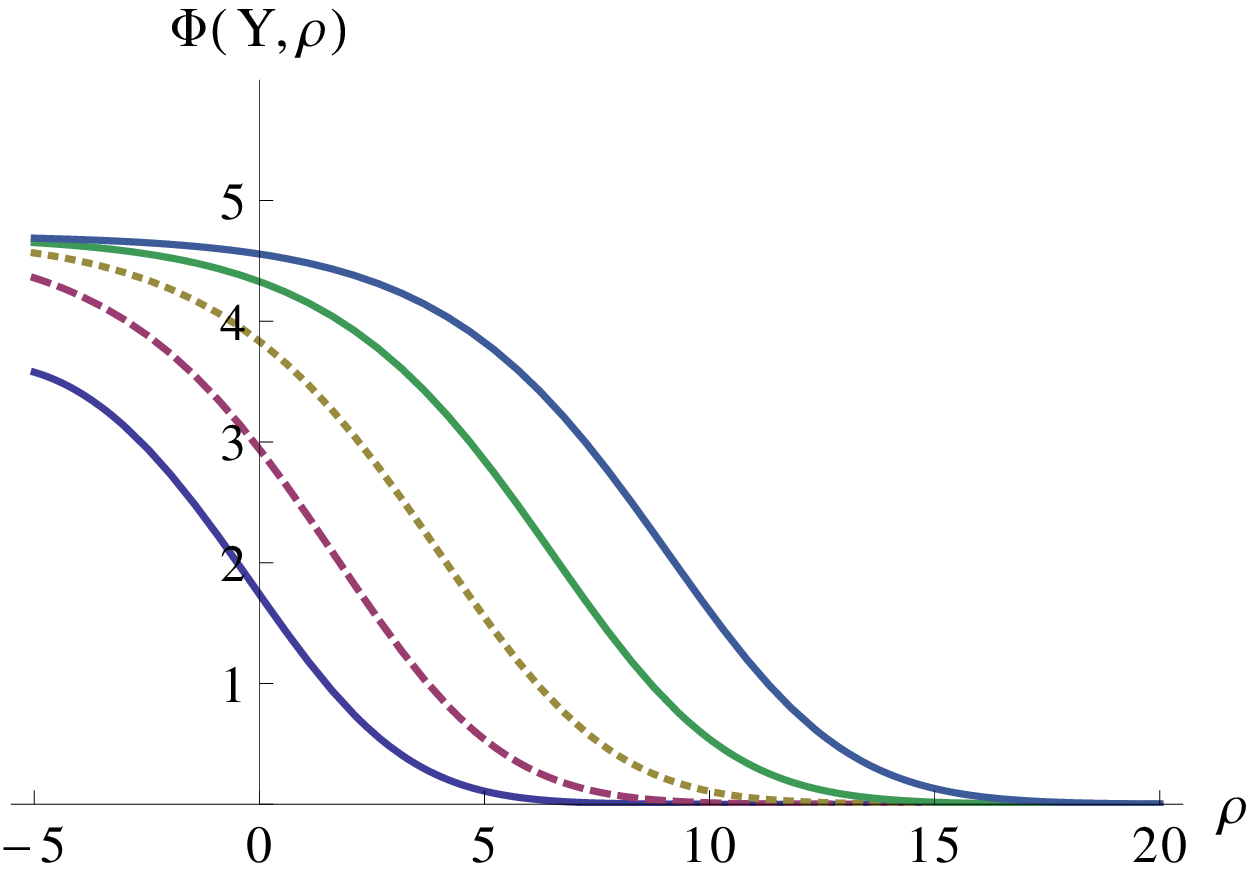}
    }

 \put(280, -80){
      \includegraphics{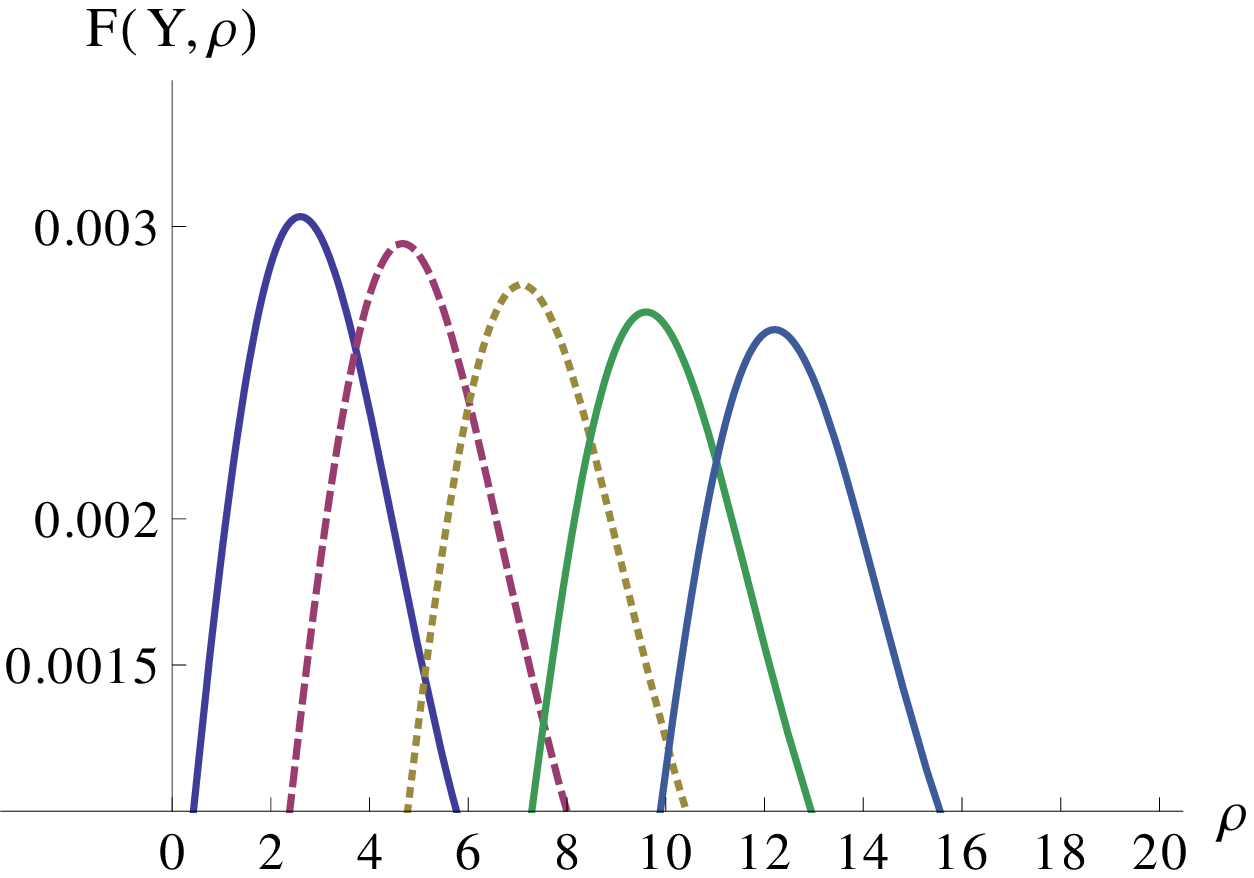}
    }

\end{picture}
\vspace{3cm}
\caption{\em \small  Left: Solution of the strongly coupled BK equation for $Y$=3, 5, 7, 9, 11. Right: The corresponding dipole gluon density. }
\vspace{1cm}
\label{fig:plotvel66}
\end{figure}
\be
\Phi_1(x,k^2)=\bar{\alpha}_s\int_{x}^1\frac{dz}{z}\left\{\int_0^\infty\frac{dl^2}{l^2}\left[\frac{l^2\Phi(x/z,l^2)-k^2\Phi(x/z,k^2)}{|l^2-k^2|}+\frac{k^2\Phi(x/z,k^2)}{\sqrt{4l^4+k^4}}\right]-\frac{\overline\alpha_s}{\pi R^2}\Phi^2(x/z,k^2)\right\}.
\ee
We note that, if one neglects the non-linear term, one recovers the linear BFKL equation. An important feature of the BK equation as observed in Ref. \cite{Munier:2003vc} is that it lies within the universality class of the Fisher-Kolmogorov-Petrovsky-Piscounov (FKPP) equation:
\be
\p _ t u(t,x)= \p ^2 _x u(t,x)+u(t,x)-u^2(t,x).
\ee
One can view this equation as a diffusion equation supplemented with a non-linear term that encodes saturation.
The question arises how to extend the BK equation to the whole strong coupling regime. We do not have an answer yet, and at this point we do not have a derivation of such an equation from first principles. Nevertheless, we can postulate such an extension based on our numerical analysis and phenomenological arguments:
\be
\partial_Y\Phi(Y,\rho)=\frac{1}{2}\lambda_{st}^{\prime} \partial^2_\rho \Phi(Y,\rho)+\frac{1}{2}\lambda_{st}^{\prime}\partial_\rho\Phi(Y,\rho)+(\lambda_{st}+\lambda_{st}^{\prime}/8)\Phi(Y,\rho)-\frac{\bar{\alpha}_s}{\pi R^2}\Phi^2(Y,\rho),
\label{eq:BKstrong}
\ee
where
the values are read off from formula (\ref{eq:fitquad}): $\lambda_{st}^{\prime}=4.08,\,\lambda_{st}=1.02$

The coefficient in front of the non-linear term has to be consistent with the large strong coupling limit we take in the linear part. We take the limit $\bar{\alpha}_s\rightarrow\infty$ ($\bar{\alpha}_s$ is essentially 't Hooft coupling) and assume a large target approximation ($R^2 \rightarrow \infty $), the ratio $\frac{\bar{\alpha}_s}{R^2}$ being fixed and we set it to unity.
The solution of the above equation with the initial condition
\be
 \Phi(Y,\rho)=e^{-\rho^2}
\ee
is presented in Fig. \ref{fig:plotvel66}, left, and it shows that at some point where the shape of the curve flattens the number of gluons saturates.  We notice that the gluon density becomes constant in the saturated regime; therefore, the derivatives vanish, and we obtain from (\ref{eq:BKstrong}) the gluon density saturation value:
\be
\Phi_{sat}=\pi(\lambda_{st}+\lambda_{st}^{\prime}/8).
\ee
\begin{figure}[t!]
  \begin{picture}(30,30)

    \put(60, -80){
      \includegraphics{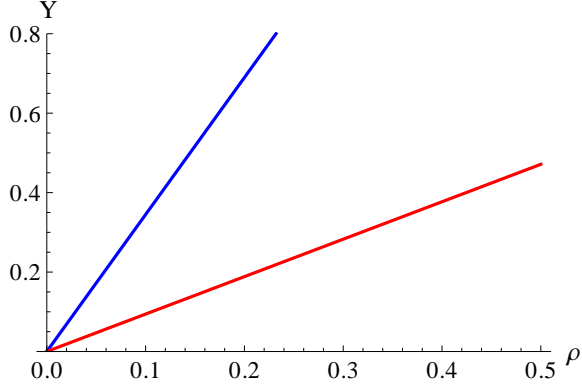}
    }

\end{picture}

\vspace{3cm}
\caption{\em {\small Red (lower) line: Saturation scale obtained from the solution of (\ref{eq:BKstrong}) using the definition (\ref{eq:satscaledef}). Blue (upper) line: saturation scale as follows from the weak coupling equations and models: $Q_s^2(Y)\simeq e^{0.29\,Y}$}}.
\label{fig:plotvel7}
\end{figure}
The behavior of the gluon number density $\Phi(Y,\rho)$ is to be contrasted with the full form of the BK in the weak coupling regime, where the rate of production of gluons slows down but still diverges logarithmically and only approximately obeys the FKKP equation.

The saturated gluons do not contribute much to the momentum distribution since the momentum gluon density or dipole gluon density which is calculated from
\be
{\cal F_{BK}}(Y,\rho)=\frac{N_c}{4\pi\alpha_s}\partial_{\rho}^2\Phi(Y,\rho)
\ee
drops off after the saturation scale has been reached (Fig. \ref{fig:plotvel66}, right) and its maxima signalize the emergence of the saturation scale which can be defined as \cite{Kutak:2009zk}
\be
\partial_{\rho} {\cal F}_{BK}(Y,\rho)|_{\rho=\ln Q_s^2(Y)}=0.
\label{eq:satscaledef}
\ee
Using the above formula, we can calculate the saturation scale that follows from our equation:
\be
Q_{s}^2(Y)\simeq e^{1.06\,Y}.
\label{eq:satline}
\ee
We plot the strong coupling saturation scale together with the weak coupling counterpart in Fig \ref{fig:plotvel7}. The above result suggests that, at equal transversal momentum, saturation effects at strong coupling occur at smaller values of $\ln (\frac{1}{x})$ than at weak coupling. This can be easily understood, since the stronger the coupling is, the closer gluons are packed, and therefore the overlapping or screening takes at the initial values of time evolution. The result (\ref{eq:satline}) is quite close to the one obtained in Refs. \cite{Hatta:2007he,Mueller:2008bt,Kovchegov:2010uk} by very different holographic methods, therefore pointing at the universality of the saturation phenomenon at large values of the coupling constant.

\section{Conclusions}

In this paper, we proposed an evolution equation that the gluon density obeys when the coupling constant is very large. We remain within a QCD framework that, through certain resummations, allows one to probe strong coupling physics. Our proposal stays within the diffusive regime, in which we investigate saturation physics. Solving this equation, we are able to extract the saturation scale, which agrees qualitatively with results from holography. We postpone the study of the running coupling effect, as well as possible extensions of the framework towards the whole range of strong coupling, for future investigations.

One very important aspect of high-energy physics is the generation of entropy after the collision. For this phenomenon we lack theoretical tools that can handle the dynamics in QCD. Therefore, it is often convenient to consider the same questions in the context of a strongly coupled plasma in the $\mathcal{N} = 4$ supersymmetric gauge theory for which one can use the AdS/CFT correspondence \cite{Heller:2011ju}. However, these methods will always be restricted to some universal properties that QCD and $\mathcal{N} = 4$ SYM share. In the context of QCD it has been suggested that the notion of a thermodynamical entropy is associated with the production of gluons in the saturation regime of dense initial states in hadron-hadron collisions \cite{Kutak:2011rb}. Later, a microscopic definition of entropy was given in Ref. \cite{Peschanski:2012cw} in which the notion of a gluon distribution function plays a crucial role. It will be interesting to employ similar ideas to study entropy generation in the model presented in this paper. Such a calculation might shed some light on the qualitative universality of the states counting in various gauge theories. We note that the entropy of the strongly coupled $\mathcal{N} = 4$ SYM theory at a
finite temperature, which is a direct measure of the number of states, differs only by a factor
3/4 from the entropy of the corresponding ideal gas, where all the states are associated with
free fundamental fields. It might be the case that QCD exhibits a similar feature. Therefore it is not unnatural that the structure of the gluon density evolution equation for both small and large values of the coupling constant is the same.


\section*{Acknowledgments}
We thank Dimitri Colferai for useful correspondence and Anna Sta\'{s}to for interesting discussions. PS acknowledges the hospitality and partial support from the Institute of Nuclear Physics, Polish Academy of Sciences where the work has been completed.\\
The work of K.K. was supported by the NCBiR Grant No. LIDER/02/35/L-2/10/NCBiR/2011. P.S. is a Postdoctoral Researcher
of FWO-Vlaanderen. The work of P.S. was supported in part by the Belgian Federal Science Policy Office through the Interuniversity Attraction
Pole IAP VI/11 and by FWO-Vlaanderen through Project No. G011410N.


\begin{appendix}
\section{Diffusion equation}\label{appdiffeq}
Following Ref. \cite{Polyanin}, we intend to solve an equation of the form
\be
\frac{\p w(t,x)}{\p t}=a \frac{\p^2 w(t,x)}{\p x^2}+b \frac{\p w(t,x)}{\p x}+c w(t,x).
\ee
The substitution
\be
w(t,x)=\exp (\beta t+ \mu x)u(t,x),
\ee
with
\be
\beta=c-\frac{b^2}{4a} \quad \mathrm{and} \quad \mu = -\frac{b}{2a},
\ee
leads to the homogenous heat equation for $u(t,x)$
\be
\frac{\p u(t,x)}{\p t}=a \frac{\p^2 u(t,x)}{\p x^2}.
\ee
As a next step we solve a Cauchy problem on the domain $-\infty\leq x \leq \infty$ with the initial condition
\be
u(0, x) = f(x).
\ee
The solution can be written as
\be
u(t,x)=\int _{-\infty} ^{\infty} d\xi f(\xi)G(t,x,\xi),
\ee
where the Green's function is
\be
G(t,x,\xi)= \frac{1}{2\sqrt{\pi a t }}\exp\left[ -\frac{(x-\xi)^2}{4at} \right].
\ee
If $f(x)=\delta(x)$, we get the solution as a Gaussian function
\be
u(t,x)=\frac{1}{2\sqrt{\pi a t }}\exp\left[ -\frac{x^2}{4at} \right].
\ee
\end{appendix}

\section{Integrals}\label{appint}
In this Appendix we collect useful formulas needed in order to evaluate the $\chi$ function.
\be
A_1=\int_0^{\infty } \frac{du}{u}\frac{u^{\gamma +\frac{\omega }{2}}}{\left|
   1-u\right| } \theta(1-u).
\ee
\be
A_2=\int_0^{\infty }\frac{du}{u}\frac{u^{\gamma -\frac{\omega }{2}}}{\left|
   1-u\right| } \theta(u-1)=\int_0^{1 }dv\frac{v^{-\gamma +\frac{\omega }{2}}}{\left|
   1-v\right| }.
\ee

\be
A_3=-\int_0^{1}du\frac{1}{u(1-u)}=-\int_0^1 du\frac{1}{1-u}-\int_0^1\frac{du}{u}.
\ee
The $A_3$ integral we split into
\be
A_{3a}=-\int_0^1 du\frac{1}{1-u},
\ee
\be
A_{3b}=-\int_0^1\frac{du}{u},
\ee
\be
A_4=-\int_1^{\infty}du\frac{1}{u(u-1)}=-\int_0^1d v\frac{1}{1-v},
\ee
\be
A_5=\int_0^{\infty}\frac{du}{u}\frac{1}{\sqrt{4u^2+1}},
\ee
\be
\int_0^1 \frac{u^{\gamma +\frac{\omega
   }{2}-1}-1}{1-u} \, du-\int_0^1 \frac{1}{u-1} \, du=\psi(1)-\psi(1-\gamma+\frac{\omega}{2})-\int_0^1 \frac{1}{u-1} \, du.
\ee
We combine the integrals in the following way:
\be
A_1+A_4=\psi(1)-\psi(\gamma+\omega/2),
\ee
\be
A_2+A_{3a}=\psi(1)-\psi(1-\gamma+\omega/2).
\ee
Introducing regulators in the remaining integrals, we get
\be
A_5+A_{3b}=
\int_0^\infty\frac{du}{u}u^{\epsilon}\frac{1}{\sqrt{4u^2+1}}-\int_0^1\frac{du}{u}u^{\epsilon}=\frac{2^{-\epsilon -1} \Gamma \left(\frac{1}{2}-\frac{\epsilon
   }{2}\right) \Gamma \left(\frac{\epsilon }{2}\right)}{\sqrt{\pi }}-\frac{1}{\epsilon}.
\ee





\end{document}